\documentclass[11pt]{article}

\usepackage[preprint]{acl}

\usepackage{times}
\usepackage{latexsym}
\usepackage[T1]{fontenc}
\usepackage[utf8]{inputenc}
\usepackage{microtype}
\usepackage{inconsolata}
\usepackage{graphicx}

\usepackage{amsmath, amssymb}
\usepackage{booktabs}
\usepackage{multirow}
\usepackage[most]{tcolorbox}
\usepackage{tabularx}
\tcbuselibrary{raster, skins, breakable}

\tcbset{
  academicbox/.style={
    enhanced,
    colback=gray!5!white,
    colframe=gray!50!black,
    boxrule=0.5pt,
    arc=2pt,
    left=6pt, right=6pt, top=4pt, bottom=4pt,
    fonttitle=\bfseries\small,
    title filled=false,
    coltitle=black,
    attach boxed title to top left={yshift=-2mm, xshift=4mm},
    boxed title style={colback=white, boxrule=0.5pt},
    breakable
  },
  configbox/.style={
    enhanced,
    colback=gray!2!white,
    colframe=gray!60!black,
    boxrule=0.5pt,
    arc=2pt,
    left=4pt, right=4pt, top=4pt, bottom=4pt,
    fonttitle=\bfseries\footnotesize,
    coltitle=white,
    colbacktitle=gray!70!black,
    attach boxed title to top left={yshift=0mm, xshift=0mm},
    boxed title style={sharp corners, boxrule=0pt},
    segmentation style={solid, draw=gray!60!black, line width=0.5pt},
    breakable
  }
}

\title{LiteRAG: Cost-Efficient Graph-Based Retrieval-Augmented Generation}

\author{\textbf{Daniel Alejandro Coll Tejeda}\quad
  \textbf{Pedro García López}\quad
  \textbf{Daniel Barcelona-Pons}\\[0.45em]
  \normalsize Universitat Rovira i Virgili\\[-0.05em]
  \small\texttt{\{danielalejandro.coll,pedro.garcia,daniel.barcelona\}@urv.cat}}

\begin{document}
\maketitle
\begin{abstract}
 Graph-based retrieval can improve multi-hop question answering, but existing approaches often incur high query-time costs and produce diffuse, oversized contexts that reduce generation efficiency. We present LiteRAG, a graph-based retrieval method that replaces expensive retrieval-time LLM control with query-conditioned algorithmic exploration and reasoning-chain context construction. On DistComp, a benchmark for multi-hop retrieval over distributed-systems papers, LiteRAG attains the highest overall quality among the evaluated methods (0.798) while reducing per-query latency by over 100$\times$ and cost by over 99\% relative to GraphRAG Global and DRIFT. On UltraDomain, it matches LinearRAG on overall quality while using about 14$\times$ fewer tokens. An ablation study indicates that LiteRAG's query-adaptive thresholding and community-aware hub penalization are the main drivers of its token-efficiency gains.

\end{abstract}

\section{Introduction}

Retrieval-Augmented Generation (RAG) has become a standard approach for grounding Large Language Models (LLMs) in external knowledge \cite{lewis2020retrieval, fan2024ragllms}. In its most common form, RAG retrieves text chunks by dense similarity and passes them to the generator as supporting context. This design is effective for many lookup-style queries, but it remains weak on questions that require combining evidence across documents, tracing relational dependencies, or synthesizing information over multiple hops \citep{gao2023retrieval, jiang2023active}.

Graph-based RAG methods address this limitation by representing corpora as structured graphs in which entities, relations, and communities are explicitly modeled \citep{edge2024graphrag, guo2024lightrag, huang2025hirag, zhuang2025linearrag}. However, many existing systems still rely heavily on the LLM at query time to traverse the graph, summarize communities, or aggregate evidence across retrieved regions. This dependence increases latency and cost, scales poorly with corpus size, and can produce broad contexts that force the generator to sift through substantial irrelevant or weakly relevant information \cite{liu2024lost}.

For multi-hop question answering, the challenge is therefore not only to retrieve enough evidence, but to present it in a form that preserves a high density of query-relevant information. When retrieval returns loosely filtered text or broad summaries, the generator must still deduce how entities connect. Under realistic token budgets, this can dilute useful evidence and reduce the value of structured retrieval.

We therefore argue that graph retrieval for RAG should optimize both evidence selection and how retrieved evidence is prepared for the generator. We introduce LiteRAG, a method that removes the generation model from the retrieval loop. Rather than relying on sequential LLM calls for navigation, LiteRAG performs query-conditioned algorithmic graph exploration: it first selects semantic and lexical anchor nodes, then expands a bounded subgraph using dynamic thresholding and community-aware hub penalization.

The final stage addresses the same objective at the context level. Rather than passing large text collections or broad summaries to the generator, LiteRAG converts the retrieved subgraph into compact reasoning chains that express the retained relations directly in the prompt. This yields a more compact, query-focused context in which relevant connections are already explicit, preserving the structure needed for generation under restricted token budgets.

Our contributions are summarized as follows:

\begin{itemize}
  \item Query-conditioned algorithmic retrieval: LiteRAG replaces fixed-hop or LLM-mediated traversal with algorithmic exploration based on dynamic thresholding and community-aware hub penalization.
  \item Reasoning-chain context construction: LiteRAG converts the retrieved subgraph into compact reasoning chains, producing a more information-dense context that remains effective within a budget of about 2,000 tokens.
  \item Quality-efficiency gains: On DistComp, LiteRAG attains the highest overall quality and lowest latency and cost among the evaluated methods; relative to the most expensive LLM-intensive GraphRAG configurations, it reduces per-query cost by over 99\% and latency by over $100\times$.
\end{itemize}

To support reproducibility, the source code for LiteRAG, benchmark queries, configurations, and evaluation scripts will be released on GitHub.

\section{Related Work}
\label{sec:related_work}

Recent work on Retrieval-Augmented Generation (RAG) \citep{fan2024ragllms} has shifted from flat passage retrieval toward structured augmentation to better support compositional multi-hop reasoning. Within this trend, graph-based methods explicitly encode entities and relations, enabling retrieval over relational structure rather than isolated text chunks.

Graph-based RAG methods can be grouped into two broad lines. A first line uses the graph to organize evidence but still relies heavily on the LLM at retrieval time. Microsoft GraphRAG \citep{edge2024graphrag} is the clearest example: LLMs are used both upstream to help construct graph representations and, more importantly, at query time to orchestrate traversal, summarize communities, and retrieve evidence from the graph. Related graph-aware reasoning systems such as G-Retriever \citep{he2024g} and Think-on-Graph \citep{sun2024think} likewise show the value of explicit relational structure. Across this line of work, the graph improves retrieval quality, but retrieval-time control remains closely coupled to the LLM.

A second line aims to make graph-based RAG more efficient. Among the most relevant systems for our setting, LightRAG \citep{guo2024lightrag} and HiRAG \citep{huang2025hirag} improve retrieval through dual-level or hierarchical organization, while related structured retrieval approaches such as RAPTOR \citep{sarthi2024raptor} reorganize evidence before generation. More algorithmic methods such as HippoRAG \citep{gutierrez2024hipporag}, LinearRAG \citep{zhuang2025linearrag}, KET-RAG \citep{huang2025ket}, and ROGRAG \citep{wang2025rograg} reduce LLM dependence through graph scoring, indexing, and filtering, but they often rely on static graph statistics or fixed heuristics and give less direct attention to constructing a dense final context for the generator.

LiteRAG belongs to this second line, but differs from prior systems in both evidence collection and final context construction. Instead of relying on fixed neighborhoods, static graph statistics, or retrieval-time LLM decisions, it performs query-conditioned subgraph construction and then converts the retained evidence into compact reasoning chains. Whereas methods such as LightRAG, HiRAG, and LinearRAG primarily improve retrieval organization or graph scoring, LiteRAG couples adaptive evidence collection with explicit construction of a more information-dense final context. Relative to Microsoft GraphRAG and related LLM-mediated systems, retrieval-time control is shifted away from the LLM.

\section{LiteRAG Method}
\label{sec:architecture}

\begin{figure*}[t]
  \centering
  \includegraphics[width=\textwidth, trim={0.6cm 0.2cm 0.6cm 0.2cm}, clip]{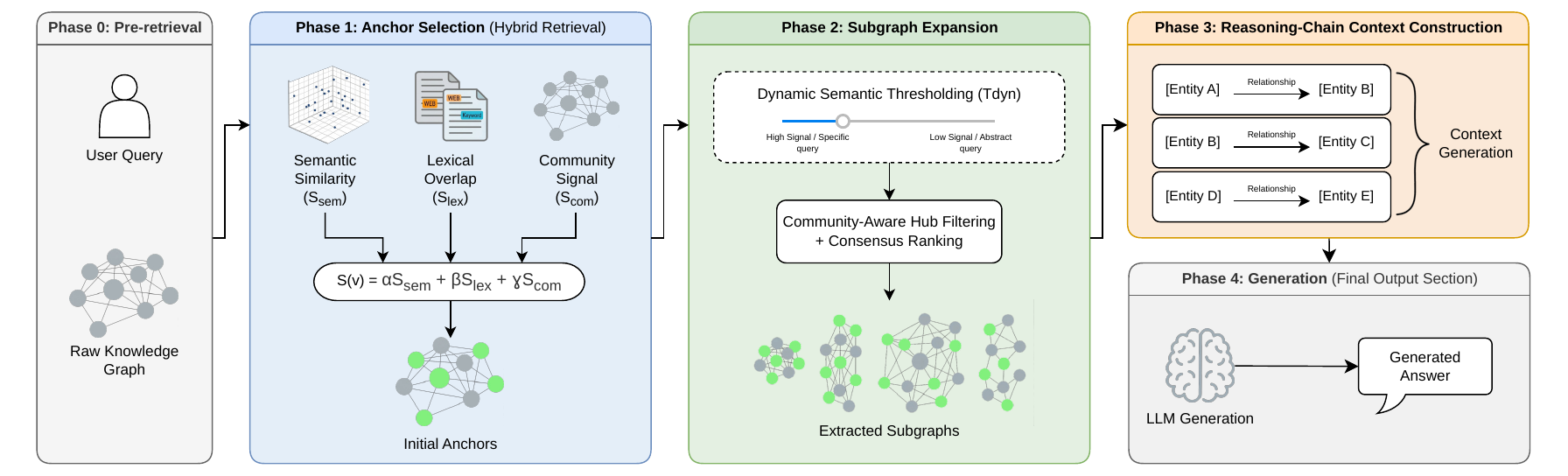}

  \caption{Overview of the LiteRAG method.}
  \label{fig:architecture}

\end{figure*}

LiteRAG is a graph-based RAG method for cost-efficient retrieval over knowledge graphs. It replaces retrieval-time LLM control with query-conditioned algorithmic exploration and explicit context construction before final answer generation. As shown in Figure~\ref{fig:architecture}, the method has three stages: (1) \textit{query-conditioned anchor selection}, which identifies promising graph entry points; (2) \textit{query-conditioned subgraph expansion}, which builds a relevant subgraph from those anchors using semantic and structural signals; and (3) \textit{reasoning-chain context construction}, which converts the retrieved subgraph into a compact representation for the generator.

\subsection{Phase 1: Query-Conditioned Anchor Selection}
\label{subsec:anchors}

Given a query $q$ and a knowledge graph $G = (V, E)$, Phase 1 selects an anchor set $A \subset V$ of semantically and lexically salient entities. Rather than scoring every node in $V$, LiteRAG retrieves a candidate pool $V_c \subset V$ via top-$K$ vector search, assigning each $v \in V_c$ a composite score $S(v)$:
\begin{equation}
  \begin{aligned}
    S(v) = {} & \alpha S_{\mathrm{sem}}(q, v) + \beta S_{\mathrm{lex}}(q, v) \\
    & {} + \gamma S_{\mathrm{com}}(q, C(v))
  \end{aligned}
  \label{eq:anchor_score}
\end{equation}
where $C(v)$ denotes the community containing $v$, and $\alpha$, $\beta$, and $\gamma$ are non-negative weights that balance the three signals. The components of the score are:
$S_{\mathrm{sem}}(q, v)$ is the semantic similarity between the query representation and node $v$, $S_{\mathrm{lex}}(q, v)$ is a lexical matching score designed to preserve entities that may be crucial by name, such as proper nouns or domain-specific terms, even when semantic similarity alone is insufficient, and $S_{\mathrm{com}}(q, C(v))$ is a community-level relevance signal that favors nodes in graph regions whose aggregate content is aligned with the query.

This formulation reflects the retrieval objective of the phase: maximize the chance that exploration starts from informative graph regions without overcommitting to a single notion of relevance. The semantic term captures conceptual alignment, the lexical term protects exact terminology, and the community term biases the search toward coherent topical neighborhoods.

After computing $S(v)$ for all candidate nodes in $V_c$, LiteRAG constructs the anchor set through thresholding:
\begin{equation}
  A = \{v \in V_c \mid S(v) \geq \tau_{\mathrm{anchor}}\}
\end{equation}
where $\tau_{\mathrm{anchor}}$ is the anchor threshold. The resulting set $A$ serves as the initial frontier for Phase 2, where LiteRAG expands a query-conditioned subgraph from these entry points.

\subsection{Phase 2: Query-Conditioned Subgraph Expansion}
\label{subsec:exploration}

Phase 2 takes the anchor set $A$ from Phase 1 and the knowledge graph $G = (V, E)$, and returns a query-conditioned subgraph $G_q = (V_q, E_q)$ expanded from those anchors. Expansion proceeds from all anchors in parallel. Rather than delegating retrieval-time navigation to an LLM, LiteRAG treats graph traversal as an algorithmic search problem in which candidate expansions are accepted or pruned according to a query-conditioned relevance score.

Concretely, LiteRAG performs a parallel subgraph expansion initialized at the nodes in $A$. At each step, the current frontier contains nodes already admitted into the explored region. For a frontier node $u$ and one of its neighbors $v$, LiteRAG computes a traversal relevance score that determines whether $v$ should be incorporated into the explored subgraph and added to the next frontier. Expansion therefore grows adaptively around the anchors, and the final output is the subgraph induced by the retained nodes and traversed edges. Exploration stops when no frontier nodes satisfy the relevance criterion below, when the maximum hop depth $k_{\max}$ is reached, or when the per-anchor expansion cap $N_{\max}$ is exhausted.

\subsubsection{Query-Adaptive Thresholding}
The first component of the exploration rule is a query-conditioned threshold that controls how selective traversal should be for a given query. Instead of using a global fixed threshold, LiteRAG derives a dynamic threshold from the quality of the initial anchors:
\begin{equation}
  \tau_{\mathrm{dyn}}(q, A) = \min\!\left(1.0,\; \tau_{\mathrm{base}} + \lambda \max_{a \in A} S(a)\right)
\end{equation}
where $\tau_{\mathrm{base}}$ is a minimum relevance floor, $\lambda$ controls how strongly anchor evidence influences the threshold, and $S(a)$ is the anchor score from Phase 1; the outer minimum caps $\tau_{\mathrm{dyn}}$ at $1.0$ because $S(a)$ is an uncalibrated composite score rather than a strict probability. This formulation links exploration directly to the quality of the retrieval starting points: strong anchors induce a stricter traversal criterion, while weaker anchors allow broader exploration of potentially relevant graph regions.

\subsubsection{Community-Aware Hub Penalization}
The second component of the exploration rule addresses high-degree hubs, which can connect otherwise unrelated graph regions and cause the search to drift. To reduce this effect, LiteRAG penalizes candidate expansions through nodes whose structural centrality is more likely to reflect generic connectivity than query relevance.

For a frontier node $u$, a neighboring candidate node $v$, and hop depth $k$, LiteRAG defines the traversal relevance as:
\begin{equation}
  R(q, u, v, k) = S_{\mathrm{sem}}(q, v) \cdot d^{k} \cdot P_{\mathrm{hub}}(u, v)
\end{equation}
where $S_{\mathrm{sem}}(q, v)$ is the semantic similarity between the query and node $v$, $d$ is a depth-decay factor, and $P_{\mathrm{hub}}(u, v)$ is a hub penalty. A candidate expansion is retained when:
\begin{equation}
  R(q, u, v, k) \geq \tau_{\mathrm{dyn}}(q, A)
\end{equation}
The hub penalty is defined as:
\begin{equation}
  P_{\mathrm{hub}}(u, v) = \frac{1}{1 + \delta \log(1 + \deg(v)) \omega(u, v)}
\end{equation}
where $\deg(v)$ is the degree of node $v$, $\delta$ controls the strength of the hub penalty, and $\omega(u, v)$ modulates this penalty according to community alignment:
\begin{equation}
  \omega(u, v) = 1 - \mathbb{I}(C(u) = C(v)) \kappa
\end{equation}
Here, $\mathbb{I}(\cdot)$ is the indicator function, $C(u)$ and $C(v)$ denote the communities of $u$ and $v$, and $\kappa \in [0,1]$ controls how strongly within-community transitions are protected from hub penalization. As a result, high-degree nodes that remain within the same topical region are penalized less aggressively than hubs that bridge unrelated communities.

Together, the dynamic threshold and hub-aware penalty define the expansion rule used to grow the frontier from the anchors. The output of this phase is the retained query-conditioned subgraph $G_q$, which is then passed to Phase 3 for ranking and context construction.

\subsection{Phase 3: Reasoning-Chain Context Construction}
\label{subsec:context}

Phase 3 converts the query-conditioned subgraph $G_q$ into the final generation context. It aims to preserve relational evidence while expressing it in a compact form under strict token budgets.

To do so, LiteRAG ranks entities in $G_q$ and retains relations incident to the top-ranked ones for inclusion in the final reasoning-chain context. For each explored entity $v \in V_q$, LiteRAG computes a consensus score $S_{\mathrm{cons}}(v)$ that combines four normalized signals:
\begin{equation}
  \begin{aligned}
    S_{\mathrm{cons}}(v) = {} & \rho_1 I_{\mathrm{rate}}(v) + \rho_2 S_{\mathrm{sem}}(q, v) \\
    & {} + \rho_3 P_{\mathrm{prox}}(v) + \rho_4 S_{\mathrm{str}}(v)
  \end{aligned}
\end{equation}
where $I_{\mathrm{rate}}(v)$ is the fraction of anchors whose expansions reach $v$, $S_{\mathrm{sem}}(q, v)$ is the semantic alignment between $v$ and the query, $P_{\mathrm{prox}}(v)$ is a topological proximity term that decays with the average hop distance from the anchors, and $S_{\mathrm{str}}(v)$ is a structural centrality score derived from graph measures such as PageRank and betweenness. To reward evidence supported by consistently relevant traversal paths, LiteRAG further rescales this score by the average traversal relevance accumulated when reaching $v$ and by a small bonus when $v$ is itself an initial anchor. Entities with the highest resulting scores are retained, and the relations connecting them are converted into a compact context representation that exposes relational structure directly to the generator.

The main representation used in this phase is a \textit{reasoning chain}, which linearizes a graph relation into a short relational statement rather than a larger unstructured text block:
\begin{quote}
  \small \texttt{Entity A} $\xrightarrow{\textit{[Relationship]}}$ \texttt{Entity B}: \textit{"Description of the specific interaction..."}
\end{quote}

The final generation context combines these reasoning chains with minimal supporting information needed for interpretation, such as definitions of top-ranked entities or brief source-grounded evidence when available. In this way, LiteRAG converts the explored subgraph into a structurally explicit context that is substantially more compact than raw-text or community summaries.

This stage is related to \textit{context compression} approaches \cite{jiang-etal-2023-llmlingua}, but its role here is more specific: rather than shortening arbitrary text, LiteRAG constructs a curated, information-dense relational context in which the relevant connections are already explicit.

Section~\ref{subsec:ablation} revisits this design choice and shows that the context-construction step reduces token consumption while preserving the explicit relational evidence needed for multi-hop reasoning.

\section{Experimental Methodology}
\label{sec:methodology}

We evaluate LiteRAG on DistComp, a custom benchmark designed to test the query types and corpus-scaling behavior central to our setting, and on the \textit{Mix} split of UltraDomain, a public benchmark evaluated under the same model setup across representative graph-based and dense-retrieval baselines.

\subsection{Datasets}

\paragraph{DistComp.} We construct \textit{DistComp} from Distributed Computing papers published between 2020 and 2022 in nine major IEEE and ACM venues. The benchmark is designed to cover query types that are less well represented in standard public datasets but central to our setting, especially multi-hop synthesis and drift-style comparison, and to test how retrieval behavior changes as corpus size grows. We create five corpus sizes, $D = \{40, 80, 160, 640, 1280\}$, with the largest split containing approximately 500,000 tokens, and annotate 160 expert-written queries in four categories: literal citation, local reasoning, global thematic synthesis, and drift-style multi-hop comparison across distant graph regions.

\paragraph{UltraDomain.} We also evaluate on the \textit{Mix} split of UltraDomain, a public multi-domain benchmark spanning agriculture, law, and healthcare.

\subsection{Baselines}

We compare LiteRAG against five baseline families: \textbf{GraphRAG Basic} as a dense-retrieval baseline over text chunks; \textbf{Microsoft GraphRAG} \citep{edge2024graphrag} in its \textit{Local}, \textit{Global}, and \textit{DRIFT} modes; \textbf{LightRAG} \citep{guo2024lightrag} in \textit{Local}, \textit{Global}, \textit{Mix}, and \textit{Hybrid} modes; \textbf{HiRAG} \citep{huang2025hirag} in \textit{Hi}, \textit{Local}, \textit{Global}, \textit{Bridge}, and \textit{No-Bridge} modes; and \textbf{LinearRAG} \citep{zhuang2025linearrag} in its default configuration.

\subsection{Implementation Details}

All methods use \texttt{gemini-2.5-flash-lite} for generation and \texttt{gemini-embedding-001} for embeddings. LiteRAG hyperparameters were selected by grid search on a validation split and are reported in Appendix~\ref{sec:appendix_hyperparameters}. For graph construction, LiteRAG reuses Microsoft GraphRAG indexing \citep{edge2024graphrag} and differs from prior methods only in retrieval and context construction.

\subsection{Evaluation Metrics}
\label{subsec:metrics}

We report one composite quality metric and three efficiency measures. Overall quality is defined as
\begin{equation}
  \begin{aligned}
    Q_{\mathrm{total}} = {} & 0.6 \cdot Q_{\mathrm{judge}} + 0.25 \cdot Q_{\mathrm{sem}} \\
    & {} + 0.15 \cdot Q_{\mathrm{lex}}
  \end{aligned}
  \label{eq:q_total}
\end{equation}
where $Q_{\mathrm{judge}}$ is an LLM-as-a-judge score, $Q_{\mathrm{sem}}$ is embedding-based semantic similarity, and $Q_{\mathrm{lex}}$ is ROUGE-L. The weights were fixed on a held-out validation split to emphasize reasoning quality while retaining semantic and lexical grounding; disaggregated values and the final weighting are reported in Appendix~\ref{sec:appendix_disaggregated_metrics}. For efficiency, we report end-to-end latency in seconds per query, total token consumption, and cost per query under a shared API pricing model (\$0.10 per 1M input tokens and \$0.40 per 1M output tokens).

\section{Results and Empirical Analysis}
\label{sec:results}

This section reports LiteRAG's results in terms of quality, latency, token usage, and cost.

\subsection{Comparative Performance on DistComp and UltraDomain}
\label{subsec:comparative_performance}

\begin{table*}[t]
  \centering
  \small
  \resizebox{\textwidth}{!}{
  \begin{tabular}{ll cccr cccr}
    \toprule
    \multirow{2}{*}{\textbf{Method}} & \multirow{2}{*}{\textbf{Configuration}} & \multicolumn{4}{c}{\textbf{DistComp $D_{1280}$}} & \multicolumn{4}{c}{\textbf{UltraDomain \textit{Mix}}} \\
    \cmidrule(lr){3-6} \cmidrule(lr){7-10}
    & & \textbf{$Q_{\mathrm{total}}$} $\uparrow$ & \textbf{Lat. (s)} $\downarrow$ & \textbf{Tokens} $\downarrow$ & \textbf{Cost (\$)} $\downarrow$ & \textbf{$Q_{\mathrm{total}}$} $\uparrow$ & \textbf{Lat. (s)} $\downarrow$ & \textbf{Tokens} $\downarrow$ & \textbf{Cost (\$)} $\downarrow$ \\
    \midrule
    \textbf{LiteRAG} & \textbf{---} & \textbf{0.798} & \textbf{1.42} & \textbf{2,291} & \textbf{0.0003} & \textbf{0.801} & \textbf{1.24} & \textbf{4,862} & \textbf{0.0006} \\
    \midrule
    GraphRAG & Basic (Naive) & 0.775 & 1.96 & 4,599 & 0.0006 & 0.778 & 38.49 & 12,835 & 0.0014 \\
    GraphRAG & Local & 0.773 & 4.27 & 7,866 & 0.0009 & 0.771 & 4.66 & 10,082 & 0.0011 \\
    GraphRAG & Global & 0.714 & 162.66 & 643,909 & 0.0679 & 0.604 & 135.34 & 519,609 & 0.0533 \\
    GraphRAG & DRIFT & 0.657 & 142.04 & 6,441,013 & 0.7901 & 0.572 & 207.66 & 33,160,141 & 3.9303 \\
    \midrule
    LightRAG & Local & 0.762 & 3.17 & 16,083 & 0.0018 & 0.789 & 7.51 & 46,309 & 0.0048 \\
    LightRAG & Global & 0.694 & 3.06 & 15,594 & 0.0017 & 0.747 & 8.45 & 38,204 & 0.0039 \\
    LightRAG & Mix & 0.760 & 3.62 & 20,361 & 0.0022 & 0.796 & 7.85 & 88,449 & 0.0090 \\
    LightRAG & Hybrid & 0.763 & 3.32 & 20,567 & 0.0022 & 0.791 & 7.59 & 69,480 & 0.0071 \\
    \midrule
    HiRAG & Hi & 0.767 & 10.95 & 29,988 & 0.0031 & 0.783 & 6.15 & 28,200 & 0.0029 \\
    HiRAG & Local & 0.777 & 8.03 & 14,296 & 0.0016 & 0.785 & 5.49 & 18,792 & 0.0020 \\
    HiRAG & Global & 0.773 & 3.17 & 20,112 & 0.0021 & 0.790 & 6.54 & 20,177 & 0.0021 \\
    HiRAG & Bridge & 0.760 & 2.68 & 16,631 & 0.0018 & 0.786 & 5.93 & 16,149 & 0.0017 \\
    HiRAG & No-Bridge & 0.754 & 2.70 & 26,432 & 0.0028 & 0.782 & 6.13 & 29,549 & 0.0031 \\
    \midrule
    LinearRAG & --- & 0.784 & 8.01 & 3,077 & 0.0006 & \textbf{0.801} & 4.79 & 68,774 & 0.0071 \\
    \bottomrule
  \end{tabular}
  }
  \caption{\label{tab:main_results} Main results comparing quality ($Q_{\mathrm{total}}$), latency (seconds/query), token consumption, and cost (\$/query) across the DistComp $D_{1280}$ benchmark and the UltraDomain \textit{Mix} split.}
\end{table*}

Table~\ref{tab:main_results} reports the main comparison on the largest DistComp setting, $D_{1280}$, and the \textit{Mix} split of UltraDomain. We report overall quality, $Q_{\mathrm{total}}$, together with the three efficiency metrics from Section~\ref{subsec:metrics}.

On $D_{1280}$, LiteRAG attains the highest overall quality ($Q_{\mathrm{total}} = 0.798$), followed by LinearRAG ($0.784$) and HiRAG Local ($0.777$). It also records the lowest latency ($1.42$s), token usage (2,291), and cost per query (\$0.0003). The efficiency difference is largest against LLM-intensive methods: GraphRAG Global requires 162.66s and 643k tokens per query, while GraphRAG DRIFT exceeds 142s and 6.4M tokens.

The comparison with LinearRAG is especially informative because both methods reduce retrieval-time LLM reliance, but they do so through different retrieval pipelines. LinearRAG relies on transformer-based query-time entity extraction and iterative graph-ranking stages, whereas LiteRAG uses bounded query-conditioned expansion and reasoning-chain context construction. On $D_{1280}$, LiteRAG improves overall quality ($0.798$ vs. $0.784$) while reducing latency from 8.01s to 1.42s, token usage from 3,077 to 2,291, and cost from \$0.0006 to \$0.0003 per query. This comparison suggests that LiteRAG can match or improve overall quality while passing a more efficient final context to the generator.

The UltraDomain results show a similar pattern. LiteRAG and LinearRAG attain the highest overall quality ($0.801$), followed by LightRAG Mix ($0.796$). LiteRAG also records the lowest latency (1.24s), token consumption (4,862), and cost (\$0.0006) among all evaluated methods.

While LinearRAG matches LiteRAG's aggregate quality on UltraDomain, it requires 4.79 seconds, 68,774 tokens, and \$0.0071 per query. Relative to LinearRAG, LiteRAG achieves a $14\times$ reduction in token volume, about a 92\% reduction in cost, and a 74\% reduction in latency. Other high-performing baselines incur even greater overhead; LightRAG Mix averages 7.85s and \$0.0090 per query, while GraphRAG DRIFT exceeds 200s and \$3.93.

Taken together, these results indicate that LiteRAG matches the quality of the strongest graph-based methods while using a much smaller execution budget. This is consistent with our central claim that a compact, structurally curated context can preserve answer quality without requiring the LLM to process large volumes of retrieved text.

\subsection{Latency and Cost Across Corpus Scale}
\label{subsec:scalability}

\begin{figure}[t]
  \centering
  \includegraphics[width=\columnwidth]{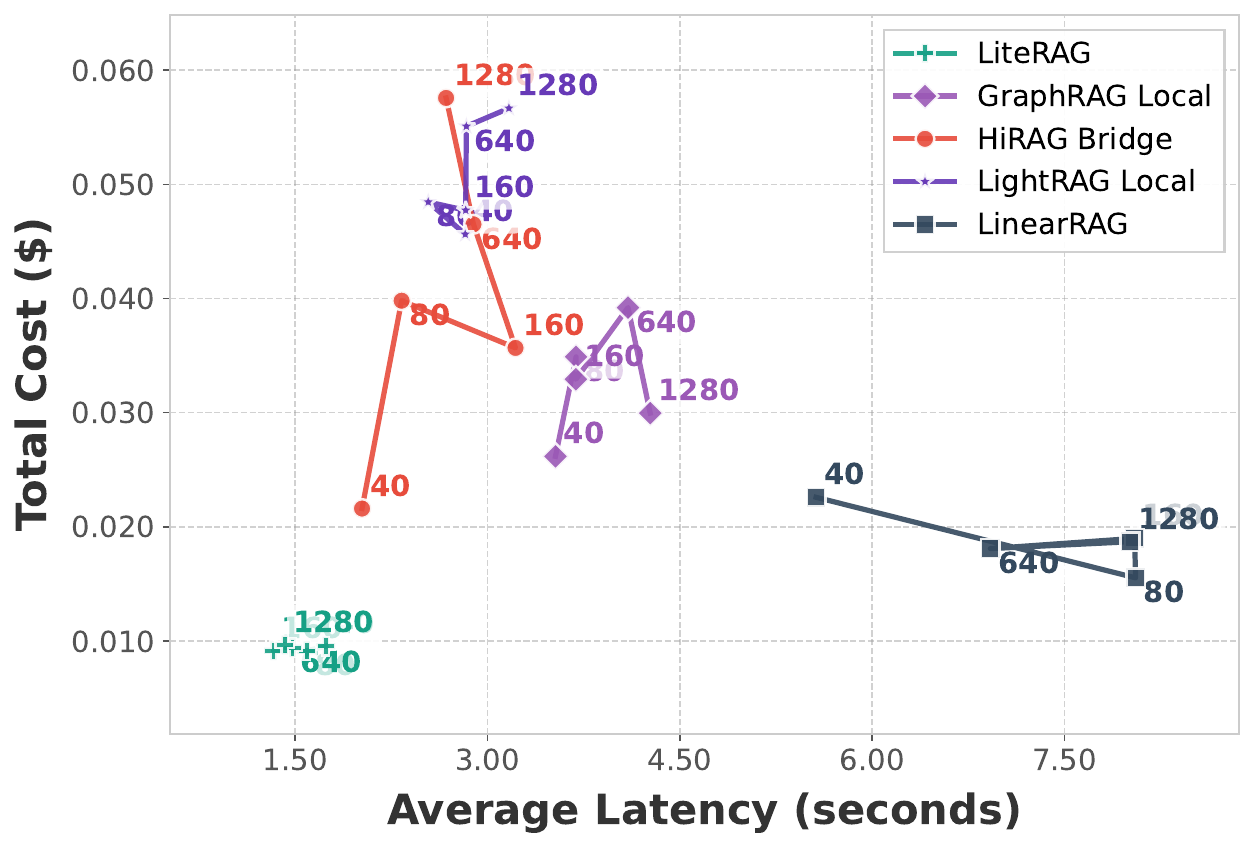}
  \caption{Latency-cost comparison across corpus sizes for LiteRAG and the best reported baseline configuration from each compared method family; labels indicate dataset size ($D_{40}$, $D_{80}$, $D_{160}$, $D_{640}$, and $D_{1280}$).}
  \label{fig:latency_scale}
\end{figure}

We next examine how efficiency changes as the DistComp corpus grows from 40 to 1280 documents. Figure~\ref{fig:latency_scale} shows LiteRAG and the strongest baseline configuration from each method family, while Appendix~\ref{sec:appendix_full_results} reports all configurations.

\paragraph{Latency Trends}
LiteRAG shows the lowest and most stable latency profile over the evaluated range, remaining between 1.33 and 1.74 seconds per query and ending at 1.42 seconds on $D_{1280}$. The selected baselines are consistently slower: LightRAG Local ranges from 2.54 to 3.17 seconds, HiRAG Bridge from 2.02 to 3.22 seconds, GraphRAG Local from 3.53 to 4.27 seconds, and LinearRAG from 5.56 to 8.05 seconds. The gap is present at every corpus size and becomes much larger for the more expensive graph-based modes reported in the appendix.

The appendix makes this separation clearer. GraphRAG Global grows from 8.57s ($D_{40}$) to 162.66s ($D_{1280}$), and DRIFT remains above 139s throughout. Over the same settings, LiteRAG stays within a narrow band. The reported results therefore show that LiteRAG preserves low end-to-end latency as corpus size increases, whereas broader graph aggregation and agentic traversal incur much larger runtime overhead.

\paragraph{Cost Trends}
The cost results follow the same pattern. LiteRAG remains between \$0.00028 and \$0.00030 per query across all corpus sizes. The selected baselines remain more expensive: LinearRAG reaches \$0.0006 on $D_{1280}$, LightRAG Local ranges from \$0.0014 to \$0.0018, HiRAG Bridge from \$0.0007 to \$0.0018, and GraphRAG Local from \$0.0008 to \$0.0012. The appendix again shows the largest increases in the LLM-intensive configurations, with GraphRAG Global rising from \$0.0031 to \$0.0679 and GraphRAG DRIFT remaining above \$0.6195 throughout.

Taken together, these results show that LiteRAG maintains the lowest reported latency-cost profile over the evaluated range. This is consistent with the design in Section~\ref{sec:architecture}: bounded candidate selection, thresholded query-conditioned expansion, and compact context construction rather than corpus-level summarization or agentic traversal. Section~\ref{subsec:token_limit} examines the same pattern under an explicit fixed-budget constraint.

\subsection{Quality Under a Fixed Token Budget}
\label{subsec:token_limit}

This experiment evaluates answer quality when each method is constrained to a context budget of about 2,000 tokens, comparable to LiteRAG's default setting. For each baseline, the budgeted variant tightens the retrieval configuration so that less context is passed to the generator while keeping the shared generation model unchanged.

\begin{table}[t]
  \centering
  \small
    \begin{tabular}{lccc}
      \toprule
      \textbf{Method} & \textbf{Default} & \textbf{Budgeted} & \textbf{$\Delta$ (\%)} \\
      \midrule
      \textit{GraphRAG} & & & \\
      Basic & 77.5\% & 72.3\% & $-5.2\%$ \\
      Local & 77.3\% & 73.3\% & $-4.0\%$ \\
      Global & 71.4\% & --- & --- \\
      DRIFT & 65.7\% & 52.4\% & $-13.3\%$ \\
      \midrule
      \textit{LightRAG} & & & \\
      Global & 69.4\% & 53.3\% & $-16.1\%$ \\
      Hybrid & 76.3\% & 71.9\% & $-4.4\%$ \\
      Local & 76.2\% & 66.9\% & $-9.3\%$ \\
      Mix & 76.0\% & 67.3\% & $-8.7\%$ \\
      \midrule
      \textit{HiRAG} & & & \\
      Hi & 76.7\% & 72.9\% & $-3.8\%$ \\
      Local & 77.7\% & 75.7\% & $-2.0\%$ \\
      Global & 77.3\% & 73.6\% & $-3.7\%$ \\
      Bridge & 76.0\% & 74.8\% & $-1.2\%$ \\
      No-Bridge & 75.4\% & 74.7\% & $-0.7\%$ \\
      \midrule
      LinearRAG & 78.4\% & 76.3\% & $-2.1\%$ \\
      \midrule
      \textbf{LiteRAG} & \textbf{79.8\%} & \textbf{79.8\%} & \textbf{0.0\%} \\
      \bottomrule
    \end{tabular}
  \caption{Default and budgeted $Q_{\mathrm{total}}$ under a fixed token budget of approximately 2,000 tokens; LiteRAG's budgeted setting is identical to its default setting.}
  \label{tab:token_limit}
\end{table}

LiteRAG is unchanged because its pipeline already meets the budget. Table~\ref{tab:token_limit} therefore shows no quality loss for LiteRAG, whereas nearly all baselines decline. The largest drops are LightRAG Global ($-16.1\%$), GraphRAG DRIFT ($-13.3\%$), LightRAG Local ($-9.3\%$), and LightRAG Mix ($-8.7\%$). GraphRAG Basic and Local drop by 5.2\% and 4.0\%, while LinearRAG declines by 2.1\%. HiRAG is the most resilient baseline family, but its strongest configurations still remain below LiteRAG's budgeted score.

GraphRAG Global could not be evaluated because its Map-Reduce retrieval procedure does not expose a comparable hard token cap. GraphRAG DRIFT is also informative: even its most restrictive available configuration still uses 570,440 tokens, down from 6,441,013, and still loses 13.3\% in quality. Some retrieval pipelines therefore cannot be reduced to a small prompt budget without either substantial degradation or budget violation.

The contrast with LiteRAG is consistent with Phase 3. LiteRAG ranks the retrieved subgraph and constructs the final reasoning-chain context from the retained evidence, so the standard retrieval output already fits the target budget without an additional compression step.

Taken together, these results show that LiteRAG's advantage is not only lower average token use, but also stronger quality retention under an explicit budget constraint. This complements the scalability analysis: the same bounded retrieval and context-construction stages that reduce latency and cost also preserve answer quality when context is tightly limited.

\subsection{Ablation Study}
\label{subsec:ablation}

To isolate the contribution of LiteRAG's main design choices, we conduct an ablation study on 32 complex multi-hop queries. We systematically disable key pipeline components and measure the resulting changes in overall quality ($Q_{\mathrm{total}}$), token consumption, and cost (Table~\ref{tab:ablation_results}).

\begin{table}[ht]
  \centering
  \small
  \setlength{\tabcolsep}{4pt}
  \begin{tabular}{lccc}
    \toprule
    \textbf{Configuration} & \textbf{$Q_{\mathrm{total}}$} $\uparrow$ & \textbf{Tokens} $\downarrow$ & \textbf{Cost (\$)} $\downarrow$ \\
    \midrule
    \textbf{LiteRAG (Full)} & \textbf{0.798} & \textbf{2,291} & \textbf{0.00030} \\
    \midrule
    Semantic-only anchors & 0.761 & 2,103 & 0.00027 \\
    Fixed-hop expansion & 0.787 & 7,979 & 0.00087 \\
    No hub penalty & 0.789 & 4,522 & 0.00052 \\
    Raw subgraph context & 0.780 & 3,622 & 0.00043 \\
    \bottomrule
  \end{tabular}
  \caption{Ablation study evaluating the impact of core architectural components on overall quality, token efficiency, and cost per query.}
  \label{tab:ablation_results}
\end{table}

\textbf{Lexical and Community Signals (Phase 1):} Removing the lexical and community signals from Phase 1 (\textit{Semantic-only anchors} in Table~\ref{tab:ablation_results}) reduces overall quality from $0.798$ to $0.761$. This is consistent with the role of query-conditioned anchor selection: semantic similarity alone is less reliable for preserving exact entity terminology and for biasing retrieval toward relevant graph regions.

\textbf{Query-Adaptive Thresholding (Phase 2):}
Replacing query-adaptive thresholding with a rigid fixed-hop expansion (\textit{Fixed-hop expansion} in Table~\ref{tab:ablation_results}) leaves overall quality relatively stable ($0.787$), but increases average token consumption by nearly 250\% (from 2,291 to 7,979) and roughly triples per-query cost. This pattern suggests that fixed-hop traversal pulls in substantially more low-yield neighboring context. Within this ablation, query-adaptive thresholding is therefore the main contributor to LiteRAG's token-efficiency.

\textbf{Community-Aware Hub Penalization (Phase 2):}
Removing community-aware hub penalization (\textit{No hub penalty} in Table~\ref{tab:ablation_results}) increases token consumption to 4,522 while leaving overall quality similar ($0.789$). This suggests that, without the penalty, high-degree hubs mainly add structural noise and cost rather than useful evidence.

\textbf{Reasoning-Chain Context Construction (Phase 3):}
Bypassing reasoning-chain context construction (\textit{Raw subgraph context} in Table~\ref{tab:ablation_results}) and instead passing the retrieved subgraph to the LLM as raw lists of entities and relationships increases token consumption by nearly $60\%$ (3,622 tokens) and lowers overall quality to $0.780$. This result is consistent with the role of Phase 3: structuring retained graph evidence into a more compact and directly usable final context.

\section{Conclusion}

In this paper, we presented LiteRAG, a graph-based retrieval method for RAG built on a simple premise: for multi-hop generation, retrieval should not only find relevant evidence, but deliver it to the LLM in a compact, structurally explicit form. LiteRAG addresses this objective by replacing LLM-mediated navigation with query-conditioned algorithmic exploration and reasoning-chain context construction. In particular, the method combines lexical and community-aware anchor selection, query-adaptive thresholding, community-aware hub penalization, and compact reasoning-chain construction before final generation.

Across the reported experiments, this design yields a favorable quality-efficiency trade-off. On DistComp $D_{1280}$, LiteRAG attains the highest reported $Q_{\mathrm{total}}$ among the compared methods while also using the fewest tokens, the lowest latency, and the lowest per-query cost. The broader scaling results show a comparatively flat latency-cost profile over the evaluated corpus range, and the fixed-budget analysis suggests that LiteRAG's advantage is not merely retrieving less context, but retrieving context with higher effective information density and remaining effective under a small budget. On UltraDomain, LiteRAG remains competitive in aggregate quality while using a substantially smaller retrieved context and markedly lower execution cost than the strongest comparator.

Taken together, these findings support the paper's three main contributions. The comparative results establish the overall quality-efficiency gains; the fixed-budget and ablation results show that reasoning-chain context construction helps preserve overall quality under tight token budgets; and the ablation study further shows that lexical and community-aware anchor selection supports retrieval quality, while query-adaptive thresholding and community-aware hub penalization are the main drivers of token-efficiency. More broadly, separating graph exploration from LLM inference and treating context construction as a core part of retrieval yields a smaller, more information-dense final context for the generator. Future work can refine retrieval controls and test the same context-curation principle across additional domains.

\section*{Limitations}

This study has some limitations. First, DistComp is a specialized benchmark centered on distributed-systems literature, so the absolute performance levels reported here should not be assumed to transfer unchanged to every domain; the UltraDomain results are intended to check that the main quality-efficiency pattern is not confined to that corpus. Second, LiteRAG focuses on retrieval and context construction rather than graph indexing, so the paper does not address the full end-to-end optimization problem when indexing cost is itself a primary concern. Third, all systems are evaluated under a shared generation and embedding setup to isolate retrieval behavior, which improves comparability but leaves cross-model robustness for future work. Finally, the scalability analysis is empirical rather than formal, and some baseline architectures do not expose controls that permit perfectly matched hard token budgets across methods. These limitations primarily affect scope and generality, rather than the paper's central claim that LiteRAG delivers a strong quality-efficiency trade-off through query-conditioned graph exploration and compact context construction.

\bibliography{custom}

\appendix
\onecolumn

\section{Complete Scalability Results}
\label{sec:appendix_full_results}

Table~\ref{tab:scalability_all_engines} reports the complete scalability results for all evaluated query engines across all dataset sizes, expanding on the summary provided in Section~\ref{subsec:scalability}. The table details average end-to-end latency, total token consumption, cost per query, and overall quality.

\begin{table*}[htbp]
  \centering
  \small
  \renewcommand{\arraystretch}{0.94}
  \setlength{\tabcolsep}{3pt}
  \begin{minipage}[t]{0.49\textwidth}
    \vspace{0pt}
    \centering
    \begin{tabular}{@{}p{1.65cm}crrrr@{}}
      \toprule
      \textbf{Engine} & \textbf{D} & \textbf{Lat. (s)} & \textbf{Tok.} & \textbf{Cost (\$)} & \textbf{$Q_{\mathrm{total}}$} \\
      \midrule
      \multirow{5}{1.65cm}{\raggedright GraphRAG Basic}
      & 40 & 2.13 & 4,417 & 0.0006 & 0.8067 \\
      & 80 & 1.62 & 4,509 & 0.0006 & 0.7938 \\
      & 160 & 1.56 & 4,354 & 0.0006 & 0.8213 \\
      & 640 & 1.84 & 4,539 & 0.0006 & 0.7906 \\
      & 1280 & 1.96 & 4,599 & 0.0006 & 0.7746 \\
      \midrule
      \multirow{5}{1.65cm}{\raggedright GraphRAG Local}
      & 40 & 3.53 & 6,482 & 0.0008 & 0.7547 \\
      & 80 & 3.69 & 9,331 & 0.0011 & 0.7818 \\
      & 160 & 3.69 & 8,739 & 0.0010 & 0.7152 \\
      & 640 & 4.10 & 10,630 & 0.0012 & 0.7610 \\
      & 1280 & 4.27 & 7,866 & 0.0009 & 0.7727 \\
      \midrule
      \multirow{5}{1.65cm}{\raggedright GraphRAG Global}
      & 40 & 8.57 & 26,633 & 0.0031 & 0.7077 \\
      & 80 & 19.16 & 53,208 & 0.0059 & 0.7498 \\
      & 160 & 30.63 & 93,364 & 0.0100 & 0.7156 \\
      & 640 & 92.68 & 348,525 & 0.0365 & 0.7088 \\
      & 1280 & 162.66 & 643,909 & 0.0679 & 0.7144 \\
      \midrule
      \multirow{5}{1.65cm}{\raggedright GraphRAG DRIFT}
      & 40 & 142.90 & 5,103,967 & 0.6470 & 0.6837 \\
      & 80 & 139.83 & 4,832,192 & 0.6195 & 0.5861 \\
      & 160 & 147.81 & 6,111,591 & 0.7502 & 0.6159 \\
      & 640 & 149.29 & 7,312,642 & 0.8777 & 0.4488 \\
      & 1280 & 142.04 & 6,441,013 & 0.7901 & 0.6574 \\
      \midrule
      \multirow{5}{1.65cm}{\raggedright LightRAG Global}
      & 40 & 3.00 & 13,178 & 0.0015 & 0.7932 \\
      & 80 & 3.08 & 13,624 & 0.0015 & 0.7657 \\
      & 160 & 2.50 & 13,859 & 0.0015 & 0.7803 \\
      & 640 & 3.09 & 15,557 & 0.0017 & 0.7524 \\
      & 1280 & 3.06 & 15,594 & 0.0017 & 0.6943 \\
      \midrule
      \multirow{5}{1.65cm}{\raggedright LightRAG Hybrid}
      & 40 & 3.52 & 17,246 & 0.0019 & 0.7939 \\
      & 80 & 3.28 & 18,491 & 0.0020 & 0.7988 \\
      & 160 & 3.34 & 18,465 & 0.0020 & 0.8103 \\
      & 640 & 3.63 & 20,256 & 0.0022 & 0.7877 \\
      & 1280 & 3.32 & 20,567 & 0.0022 & 0.7627 \\
      \midrule
      \multirow{5}{1.65cm}{\raggedright LightRAG Local}
      & 40 & 2.83 & 12,509 & 0.0014 & 0.7956 \\
      & 80 & 2.54 & 13,650 & 0.0015 & 0.7969 \\
      & 160 & 2.83 & 13,364 & 0.0015 & 0.8179 \\
      & 640 & 2.83 & 15,566 & 0.0017 & 0.8112 \\
      & 1280 & 3.17 & 16,083 & 0.0018 & 0.7622 \\
      \midrule
      \multirow{5}{1.65cm}{\raggedright LightRAG Mix}
      & 40 & 3.71 & 17,126 & 0.0019 & 0.8010 \\
      & 80 & 3.39 & 18,292 & 0.0020 & 0.7991 \\
      & 160 & 3.09 & 18,573 & 0.0020 & 0.8044 \\
      & 640 & 3.70 & 20,166 & 0.0022 & 0.7895 \\
      & 1280 & 3.62 & 20,361 & 0.0022 & 0.7596 \\
      \bottomrule
    \end{tabular}
  \end{minipage}
  \hfill
  \begin{minipage}[t]{0.49\textwidth}
    \vspace{0pt}
    \centering
    \begin{tabular}{@{}p{1.65cm}crrrr@{}}
      \toprule
      \textbf{Engine} & \textbf{D} & \textbf{Lat. (s)} & \textbf{Tok.} & \textbf{Cost (\$)} & \textbf{$Q_{\mathrm{total}}$} \\
      \midrule
      \multirow{5}{1.65cm}{\raggedright HiRAG Hi}
      & 40 & 5.92 & 11,319 & 0.0013 & 0.7761 \\
      & 80 & 2.32 & 23,914 & 0.0025 & 0.7706 \\
      & 160 & 3.92 & 22,453 & 0.0024 & 0.8049 \\
      & 640 & 2.99 & 26,958 & 0.0028 & 0.7841 \\
      & 1280 & 10.95 & 29,988 & 0.0031 & 0.7668 \\
      \midrule
      \multirow{5}{1.65cm}{\raggedright HiRAG Local}
      & 40 & 2.26 & 6,640 & 0.0008 & 0.7680 \\
      & 80 & 1.96 & 8,833 & 0.0010 & 0.7860 \\
      & 160 & 8.77 & 8,029 & 0.0009 & 0.8115 \\
      & 640 & 8.24 & 10,834 & 0.0012 & 0.7924 \\
      & 1280 & 8.03 & 14,296 & 0.0016 & 0.7768 \\
      \midrule
      \multirow{5}{1.65cm}{\raggedright HiRAG Global}
      & 40 & 2.39 & 7,764 & 0.0009 & 0.7630 \\
      & 80 & 2.16 & 15,935 & 0.0017 & 0.8069 \\
      & 160 & 8.06 & 15,164 & 0.0016 & 0.8059 \\
      & 640 & 5.35 & 17,540 & 0.0019 & 0.7905 \\
      & 1280 & 3.17 & 20,112 & 0.0021 & 0.7731 \\
      \midrule
      \multirow{5}{1.65cm}{\raggedright HiRAG Bridge}
      & 40 & 2.02 & 5,584 & 0.0007 & 0.7210 \\
      & 80 & 2.33 & 11,362 & 0.0012 & 0.7780 \\
      & 160 & 3.22 & 10,035 & 0.0011 & 0.8047 \\
      & 640 & 2.89 & 13,245 & 0.0015 & 0.7812 \\
      & 1280 & 2.68 & 16,631 & 0.0018 & 0.7602 \\
      \midrule
      \multirow{5}{1.65cm}{\raggedright HiRAG No-Bridge}
      & 40 & 2.35 & 11,330 & 0.0013 & 0.7760 \\
      & 80 & 2.23 & 20,305 & 0.0022 & 0.7915 \\
      & 160 & 4.43 & 19,425 & 0.0021 & 0.8050 \\
      & 640 & 3.52 & 22,890 & 0.0024 & 0.7788 \\
      & 1280 & 2.70 & 26,432 & 0.0028 & 0.7541 \\
      \midrule
      \multirow{5}{1.65cm}{\raggedright LinearRAG}
      & 40 & 5.56 & 3,266 & 0.0007 & 0.8081 \\
      & 80 & 8.05 & 2,649 & 0.0005 & 0.7826 \\
      & 160 & 8.05 & 2,800 & 0.0006 & 0.8223 \\
      & 640 & 6.92 & 2,904 & 0.0006 & 0.7484 \\
      & 1280 & 8.01 & 3,077 & 0.0006 & 0.7841 \\
      \midrule
      \multirow{5}{1.65cm}{\raggedright \textbf{LiteRAG}}
      & 40 & 1.74 & 2,158 & 0.0003 & 0.8090 \\
      & 80 & 1.59 & 2,195 & 0.0003 & 0.8277 \\
      & 160 & 1.33 & 2,170 & 0.0003 & 0.8398 \\
      & 640 & 1.48 & 2,235 & 0.0003 & 0.8090 \\
      & \textbf{1280} & \textbf{1.42} & \textbf{2,291} & \textbf{0.0003} & \textbf{0.7980} \\
      \bottomrule
    \end{tabular}
  \end{minipage}
  \caption{Complete scalability results for all query engines across all DistComp dataset sizes ($D$), detailing end-to-end latency, token consumption, cost per query, and $Q_{\mathrm{total}}$.}
  \label{tab:scalability_all_engines}
\end{table*}

\section{Disaggregated Quality Metrics}
\label{sec:appendix_disaggregated_metrics}

This appendix reports the disaggregated components of the composite quality metric $Q_{\mathrm{total}}$ for the top-performing configurations on the 1280-document corpus.

The composite metric is defined in Equation~\ref{eq:q_total}, with weights $0.6$, $0.25$, and $0.15$ assigned to the judge-based, semantic, and lexical components, respectively (as introduced in Section~\ref{subsec:metrics}). The weights were fixed on a held-out validation split after comparing candidate mixtures of judge-based, semantic, and lexical signals. The largest weight is assigned to $Q_{\mathrm{judge}}$ because it is the only component that directly evaluates multi-hop reasoning quality, factual correctness, and question relevance in an integrated way. In our implementation, $Q_{\mathrm{judge}}$ is normalized to the $[0,1]$ range from the aggregate scores of Correctness, Completeness, and Relevance. $Q_{\mathrm{sem}}$ provides a softer semantic alignment signal, while $Q_{\mathrm{lex}}$ (ROUGE-L) acts as a stricter lexical grounding term that penalizes severe wording-level drift or unsupported terminology.

Table~\ref{tab:disaggregated_metrics} displays the raw scores for each sub-metric. LiteRAG demonstrates consistently high performance across all three independent evaluators, suggesting that its efficiency (detailed in Section~\ref{subsec:comparative_performance}) does not come at the cost of semantic or factual degradation.

\begin{table}[htbp]
  \centering
  \small
  \begin{tabular}{lcccc}
    \toprule
    \textbf{System (Mode)} & \textbf{$Q_{\mathrm{judge}}$} & \textbf{$Q_{\mathrm{sem}}$} & \textbf{$Q_{\mathrm{lex}}$} & \textbf{$Q_{\mathrm{total}}$} \\
    \midrule
    LiteRAG (Default) & 0.909 & 0.866 & 0.242 & \textbf{0.798} \\
    GraphRAG (Basic) & 0.904 & 0.857 & 0.119 & 0.775 \\
    GraphRAG (Drift) & 0.732 & 0.807 & 0.111 & 0.657 \\
    GraphRAG (Global) & 0.822 & 0.827 & 0.095 & 0.714 \\
    GraphRAG (Local) & 0.912 & 0.837 & 0.108 & 0.773 \\
    LightRAG (Global) & 0.774 & 0.827 & 0.155 & 0.694 \\
    LightRAG (Hybrid) & 0.883 & 0.850 & 0.138 & 0.763 \\
    LightRAG (Local) & 0.879 & 0.852 & 0.144 & 0.762 \\
    LightRAG (Mix) & 0.877 & 0.850 & 0.137 & 0.760 \\
    HiRAG (Bridge) & 0.881 & 0.852 & 0.124 & 0.760 \\
    HiRAG (Global) & 0.907 & 0.854 & 0.102 & 0.773 \\
    HiRAG (Hi) & 0.894 & 0.852 & 0.117 & 0.767 \\
    HiRAG (Local) & 0.912 & 0.852 & 0.113 & 0.777 \\
    HiRAG (No-Bridge) & 0.875 & 0.849 & 0.110 & 0.754 \\
    LinearRAG (Default) & 0.851 & \textbf{0.875} & \textbf{0.365} & 0.784 \\
    \bottomrule
  \end{tabular}
  \caption{Disaggregated scores for LLM-as-a-judge ($Q_{\mathrm{judge}}$), semantic similarity ($Q_{\mathrm{sem}}$), lexical overlap ($Q_{\mathrm{lex}}$), and the composite metric ($Q_{\mathrm{total}}$) on the $D_{1280}$ benchmark.}
  \label{tab:disaggregated_metrics}
\end{table}

\section{Prompt Templates}
\label{sec:appendix_prompts}

To ensure reproducibility, we provide the exact prompt templates used for context assembly, final answer generation, and the LLM-as-a-judge evaluation. These templates were extracted directly from the system's source code.

\subsection{LiteRAG Context Assembly and Generation Prompts}
As described in Section~\ref{subsec:context}, LiteRAG avoids dumping raw text by converting graph relationships into highly dense, structured sections. During the final response generation phase, the system prompt and user prompt are constructed as follows:

\begin{tcolorbox}[academicbox, title=System Prompt]
  \small\ttfamily
  You are a helpful assistant that answers questions based on provided knowledge graph context. Be accurate, cite specific entities when relevant, and acknowledge if information is incomplete.
\end{tcolorbox}

\begin{tcolorbox}[academicbox, title=User Prompt]
  \small\ttfamily
  \#\# Knowledge Graph Context\\[0.2cm]
  {[}Direct Evidence (Source Text){]}\\
  {[}Graph Context{]}: ...\\[0.2cm]
  {[}Graph Reasoning Chains (Connections){]}\\
  $\bullet$ **Entity A** is connected to **Entity B** \\
  \hspace*{0.5cm}via *Description of the relationship*\\
  ...\\[0.2cm]
  {[}Entity Definitions{]}\\
  **Entity A**: Description of Entity A...\\
  ...\\[0.2cm]
  \#\# Question\\
  \{user\_query\}\\[0.2cm]
  \#\# Answer\\
  Based on the knowledge graph context above, provide a clear and accurate answer:
\end{tcolorbox}

By pre-computing the logical connections in the \textit{Graph Reasoning Chains} section, the prompt makes relationships explicit and reduces the need for the generator to infer them from dispersed passages, which lowers token consumption and the inferential burden on the generator.

\subsection{LLM-as-a-judge Evaluation Prompt}
For the judge-based evaluation ($Q_{\mathrm{judge}}$), the independent evaluator utilizes a detailed, criteria-based prompt designed to enforce rigorous, step-by-step reasoning. The exact prompt used in the benchmarking framework is as follows:

\begin{tcolorbox}[academicbox, title=Evaluation Prompt]
  \footnotesize\ttfamily
  You are an impartial and scientific evaluator. Your task is to assess a generated answer against a ground truth reference, based on a specific question. Evaluate the answer based on the criteria of Correctness, Completeness, and Relevance.\\[0.2cm]
  **Evaluation Task:**\\[0.2cm]
  1. Analyze the Question: Understand what the user is asking for.\\
  \hspace*{0.5cm}- Question: "\{question\}"\\[0.2cm]
  2. Analyze the Ground Truth: This is the reference for what a correct and complete answer should contain.\\
  \hspace*{0.5cm}- Ground Truth: "\{ground\_truth\}"\\[0.2cm]
  3. Analyze the Generated Answer: This is the answer you must evaluate.\\
  \hspace*{0.5cm}- Generated Answer: "\{prediction\}"\\[0.2cm]
  4. Perform a Step-by-Step Evaluation:\\[0.1cm]
  \hspace*{0.5cm}- Correctness (0.0-1.0): Is the information in the Generated Answer factually accurate and consistent with the Ground Truth? Does it contradict the ground truth or introduce plausible but unsupported information (hallucinations)?\\[0.1cm]
  \hspace*{0.5cm}- Completeness (0.0-1.0): Does the Generated Answer cover all the key information and essential points present in the Ground Truth?\\[0.1cm]
  \hspace*{0.5cm}- Relevance (0.0-1.0): Does the Generated Answer directly address the user's Question? Is the answer on-topic?\\[0.2cm]
  5. Provide Scores and Reasoning: Based on your analysis, provide a score from 0.0 (terrible) to 1.0 (perfect) for each criterion and write a detailed reasoning for your scores.\\[0.2cm]
  Return your evaluation as JSON with this exact format:\\
  \{\\
    \hspace*{0.5cm}"correctness": $<$float 0-1$>$,\\
    \hspace*{0.5cm}"completeness": $<$float 0-1$>$,\\
    \hspace*{0.5cm}"relevance": $<$float 0-1$>$,\\
    \hspace*{0.5cm}"reasoning": "$<$string$>$"\\
  \}
\end{tcolorbox}

\section{LiteRAG Hyperparameter Configuration}
\label{sec:appendix_hyperparameters}

This appendix reports the exact LiteRAG configuration used in all experiments. As noted in Section~\ref{sec:methodology}, hyperparameters were selected by grid search on a held-out validation split under the same model and graph-construction setup used for the reported results.

Table~\ref{tab:hyperparameters} instantiates the parameters introduced in the LiteRAG method (Section~\ref{sec:architecture}). For Phase 1 (Section~\ref{subsec:anchors}), the lexical term $S_{\mathrm{lex}}$ is implemented as a combination of exact and fuzzy matching, so the table decomposes the single lexical weight $\beta$ in Equation~\ref{eq:anchor_score} into its implementation-level components. For Phase 2 (Section~\ref{subsec:exploration}), the reported runs used an explicit hop-depth budget $k_{\max}=3$ together with a per-anchor expansion cap $N_{\max}=50$, so exploration was governed by the dynamic thresholding rule under bounded traversal budgets.

\begin{table*}[htbp]
  \centering
  \small
  \begin{tabularx}{\textwidth}{@{}llcX@{}}
    \toprule
    \textbf{Phase} & \textbf{Parameter} & \textbf{Value} & \textbf{Description} \\
    \midrule
    \multirow{7}{*}{\shortstack[l]{Anchor\\Selection}}
    & \texttt{candidate\_pool\_size} ($K$) & $8$ & Number of top vector-retrieved candidates retained before anchor scoring. This keeps Phase 1 local to a small semantically relevant pool rather than scoring the full graph.\\
    & \texttt{min\_anchor\_score} ($\tau_{\mathrm{anchor}}$) & $0.30$ & Minimum composite anchor score required for a node to enter $A$. The selected value preserves multiple plausible entry points without flooding Phase 2 with weak anchors.\\
    & \texttt{semantic\_weight} ($\alpha$) & $0.40$ & Contribution of semantic similarity to the Phase 1 anchor score.\\
    & \texttt{lexical\_weight} ($\beta$) & $0.45$ & Total contribution of lexical evidence to the Phase 1 anchor score. This relatively large weight protects exact terminology and named entities that may be poorly captured by embeddings alone.\\
    & \texttt{keyword\_exact\_weight} & $0.30$ & Exact-match component of $S_{\mathrm{lex}}$. This is the dominant share of $\beta$ and rewards direct terminology overlap.\\
    & \texttt{keyword\_fuzzy\_weight} & $0.15$ & Fuzzy-match component of $S_{\mathrm{lex}}$, used to recover near matches and minor surface-form variations without overpowering exact lexical evidence.\\
    & \texttt{community\_weight} ($\gamma$) & $0.15$ & Contribution of community-level relevance to the Phase 1 anchor score. The smaller weight keeps community evidence supportive rather than dominant.\\
    \midrule
    \multirow{7}{*}{\shortstack[l]{Subgraph\\Expansion}}
    & \texttt{max\_exploration\_depth} ($k_{\max}$) & $3$ & Maximum hop distance from each anchor during Phase 2. This bounds query-time search radius and is the primary explicit exploration budget used in the reported runs.\\
    & \texttt{max\_expansions\_per\_anchor} ($N_{\max}$) & 50 & A per-anchor expansion cap of 50 nodes was used to prevent unbounded exploration from highly connected anchors.\\
    & \texttt{min\_relevance\_threshold} ($\tau_{\mathrm{base}}$) & $0.25$ & Base floor in the dynamic threshold $\tau_{\mathrm{dyn}}(q,A)$. This prevents traversal from expanding on very weak evidence even when anchors are modest.\\
    & \texttt{signal\_amplification} ($\lambda$) & $0.25$ & Scaling factor that raises the traversal threshold when the initial anchors are strong. The selected value makes expansion more selective for well-grounded queries while still allowing broader search when anchor quality is weaker.\\
    & \texttt{relevance\_decay\_factor} ($d$) & $0.70$ & Multiplicative decay applied per hop in $R(q,u,v,k)$. This encourages shorter explanatory paths while still permitting multi-hop retrieval.\\
    & \texttt{degree\_influence} ($\delta$) & $0.05$ & Strength of the hub penalty. The small value suppresses generic hubs without over-penalizing structurally important nodes.\\
    & \texttt{community\_cohesion} ($\kappa$) & $0.80$ & Within-community protection factor in $\omega(u,v)$. This substantially relaxes hub penalization for transitions that remain inside the same topical region.\\
    \midrule
    \multirow{4}{*}{\shortstack[l]{Consensus\\Ranking}}
    & \texttt{intersection\_weight} ($\rho_1$) & $0.30$ & Weight on $I_{\mathrm{rate}}(v)$, rewarding entities reached by multiple anchor expansions. This is the largest single Phase 3 weight because repeated recovery across paths is treated as the strongest indicator of relevance.\\
    & \texttt{semantic\_weight} ($\rho_2$) & $0.25$ & Weight on semantic alignment in the final ranking. This keeps the retained context tightly tied to the user query after graph expansion.\\
    & \texttt{proximity\_weight} ($\rho_3$) & $0.25$ & Weight on topological proximity to the anchor set. Matching the semantic term, it favors nearby evidence without excluding informative multi-hop entities.\\
    & \texttt{structural\_weight} ($\rho_4$) & $0.20$ & Weight on structural centrality. This is the smallest Phase 3 weight so that generic graph prominence does not override query-conditioned evidence.\\
    \bottomrule
  \end{tabularx}
  \caption{Exact LiteRAG hyperparameters used in the reported experiments. The table instantiates the abstract parameters from Section~\ref{sec:architecture} and records the implementation-level decomposition of the lexical anchor term.}
  \label{tab:hyperparameters}
\end{table*}

These settings were used for every LiteRAG result reported in the main paper. Together, they reflect the intended retrieval bias of the method: preserve exact lexical anchors when necessary, expand conservatively from strong initial evidence, and rank Phase 3 (Section~\ref{subsec:context}) entities primarily by repeated support across anchor-induced traversals rather than by generic graph centrality alone.

\section{Detailed Engine Configurations}
\label{sec:appendix_configurations}

This appendix provides the full configuration parameters for all retrieval engines evaluated in this study. To ensure readability and clear side-by-side comparison, we present the parameter configurations using standard academic tables. We provide both the standard settings used for the primary benchmarking (Table~\ref{tab:main_results}) and the budgeted settings used for the fixed-budget analysis (Section~\ref{subsec:token_limit}).

The default configurations used in this study were obtained from the standard parameters in the repositories of the different architectures.

\subsection{LiteRAG Configuration}
The LiteRAG configuration and hyperparameters are defined in Appendix~\ref{sec:appendix_hyperparameters}. LiteRAG does not require a separate budgeted configuration because its algorithmic design naturally produces a context within the 2,000-token target budget.

\subsection{LinearRAG Configuration}
\label{subsec:linearrag_config}
The configuration for LinearRAG follows the suggested defaults for document-based retrieval as per its reference implementation.

\begin{table}[htbp]
  \centering
  \small
  \begin{tabular}{lcc}
    \toprule
    \textbf{Parameter} & \textbf{Standard} & \textbf{Budgeted} \\
    \midrule
    \texttt{spacy\_model} & \multicolumn{2}{c}{\texttt{"en\_core\_web\_trf"}} \\
    \texttt{max\_workers} & 4 & 4 \\
    \texttt{retrieval\_top\_k} & 5 & 3 \\
    \texttt{max\_iterations} & 3 & 2 \\
    \texttt{top\_k\_sentence} & 1 & 1 \\
    \texttt{passage\_ratio} & 1.5 & 1.5 \\
    \texttt{iteration\_threshold} & 0.5 & 0.5 \\
    \texttt{use\_vectorized} & \texttt{false} & \texttt{false} \\
    \bottomrule
  \end{tabular}
  \caption{LinearRAG configuration comparison.}
\end{table}

\subsection{LightRAG Configuration}
To test LightRAG under a fixed budget, we reduced the \texttt{max\_total\_tokens} and associated entity/relation limits.

\begin{table}[htbp]
  \centering
  \small
  \begin{tabular}{lcc}
    \toprule
    \textbf{Parameter} & \textbf{Standard} & \textbf{Budgeted} \\
    \midrule
    \texttt{top\_k} & 60 & 60 \\
    \texttt{max\_entity\_tokens} & 6000 & 600 \\
    \texttt{max\_relation\_tokens} & 8000 & 600 \\
    \texttt{max\_total\_tokens} & 30000 & 2000 \\
    \texttt{enable\_rerank} & \texttt{false} & \texttt{false} \\
    \bottomrule
  \end{tabular}
  \caption{LightRAG configuration comparison.}
\end{table}

\subsection{HiRAG Configuration}
For the budgeted version, we enforced \texttt{limit\_tokens: true} and capped the total context at 2,000 tokens.

\begin{table}[htbp]
  \centering
  \small
  \begin{tabular}{lcc}
    \toprule
    \textbf{Parameter} & \textbf{Standard} & \textbf{Budgeted} \\
    \midrule
    \texttt{top\_k} & 20 & 20 \\
    \texttt{top\_m} & 10 & 10 \\
    \texttt{max\_token\_text\_unit} & 20000 & --- \\
    \texttt{max\_token\_local} & 20000 & --- \\
    \texttt{max\_token\_bridge} & 12500 & --- \\
    \texttt{max\_token\_report} & 12500 & --- \\
    \texttt{max\_total\_tokens} & --- & 2000 \\
    \texttt{limit\_tokens} & \texttt{false} & \texttt{true} \\
    \bottomrule
  \end{tabular}
  \caption{HiRAG configuration comparison.}
\end{table}

\subsection{Microsoft GraphRAG Summary}
The budgeted version capped \texttt{max\_context\_tokens} at 4,000 (the lowest functional setting for DRIFT) and reduced the basic search \texttt{k}.

\begin{table}[htbp]
  \centering
  \small
  \begin{tabular}{lcc}
    \toprule
    \textbf{Mode / Parameter} & \textbf{Standard} & \textbf{Budgeted} \\
    \midrule
    \textit{Local Search} \\
    \texttt{max\_context\_tokens} & 12000 & 4000 \\
    \addlinespace
    \textit{Global Search} \\
    \texttt{max\_context\_tokens} & 12000 & 4000 \\
    \addlinespace
    \textit{DRIFT Search} \\
    \texttt{n\_depth} & 2 & --- \\
    \texttt{concurrency} & 32 & --- \\
    \texttt{loc\_search\_max\_data\_toks} & --- & 4000 \\
    \addlinespace
    \textit{Basic Search} \\
    \texttt{k} & --- & 4 \\
    \texttt{max\_context\_tokens} & --- & 4000 \\
    \bottomrule
  \end{tabular}
  \caption{GraphRAG configuration comparison.}
\end{table}

\section{Exploratory Multi-Entity Efficiency Comparison}
\label{sec:appendix_entity_efficiency}

As an exploratory analysis extending the findings in Section~\ref{sec:results}, this appendix reports a focused comparison between LiteRAG and LinearRAG on a small DistComp subset designed to probe multi-entity queries. The subset contains 20 queries, with 5 queries in each of four bins defined by the number of distinct entities or concepts explicitly mentioned in the question. The purpose of this analysis is focused: to examine whether the efficiency gap between the two methods remains visible as the number of entities in the query increases.

\begin{table}[htbp]
  \centering
  \small
  \renewcommand{\arraystretch}{1.1}
  \begin{tabular}{ l r r r r r r }
    \toprule
    \textbf{Ent.} & \textbf{L-Lat.} & \textbf{R-Lat.} & \textbf{L-Tok.} & \textbf{R-Tok.} & \textbf{L-Cost} & \textbf{R-Cost} \\
    \midrule
    1 & 1.02 & 2.92 & 2,009 & 2,365 & 0.00022 & 0.00033 \\
    2 & 1.11 & 2.59 & 2,120 & 3,077 & 0.00024 & 0.00044 \\
    3 & 1.00 & 2.12 & 1,881 & 2,475 & 0.00020 & 0.00034 \\
    4 & 1.09 & 3.77 & 1,961 & 2,833 & 0.00022 & 0.00048 \\
    \bottomrule
  \end{tabular}
  \caption{Efficiency comparison on the 20-query multi-entity subset. \textbf{L} denotes LiteRAG and \textbf{R} denotes LinearRAG. Latency is reported in seconds per query and cost in dollars per query.}
  \label{tab:appendix_entity_efficiency}
\end{table}

Across all four bins in Table~\ref{tab:appendix_entity_efficiency}, LiteRAG remains faster, less expensive, and more token-efficient than LinearRAG. LiteRAG stays between 1.00 and 1.11 seconds and between 1,881 and 2,120 tokens, whereas LinearRAG ranges from 2.12 to 3.77 seconds and from 2,365 to 3,077 tokens. The quality differences on this subset are mixed in the lower-entity bins, while LiteRAG records the higher score on the 4-entity bin (0.81 vs. 0.74). Accordingly, the clearest conclusion from this subset is that LiteRAG preserves a more stable efficiency profile across query-complexity bins.

\section{Example Queries by Entity Complexity}
\label{sec:appendix_example_queries}

This appendix provides representative examples from the multi-entity subset summarized in Appendix~\ref{sec:appendix_entity_efficiency}. The queries are grouped by the number of distinct entities that must be related to produce a correct answer.

\paragraph{1 Entity (Simple Lookup)}
\begin{itemize}
    \item \textit{``What specific hardware mechanism does 'Poseidon' use to protect heap metadata?''}
    \item \textit{``What is the primary function of the 'IoTwins' platform?''}
\end{itemize}

\paragraph{2 Entities (Relational Reasoning)}
\begin{itemize}
    \item \textit{``How does 'SparkFlow' utilize 'Base Recalibrator' in its design?''}
    \item \textit{``In the context of binary hardening, what capability did 'RedFat' demonstrate regarding 'Google Chrome'?''}
\end{itemize}

\paragraph{3 Entities (Multi-hop Synthesis)}
\begin{itemize}
    \item \textit{``What is the relationship between 'HopsFS-CL', 'HopsFS', and 'HDFS'?''}
    \item \textit{``How does 'Siren' improve upon current 'Byzantine-robust' aggregation rules in 'Federated Learning'?''}
\end{itemize}

\paragraph{4 Entities (Comparative Analysis)}
\begin{itemize}
    \item \textit{``In the scalability evaluation of deep learning frameworks, which framework among 'Tensorflow', 'Keras', 'MXNet', and 'PyTorch' was found to be the most efficient?''}
    \item \textit{``How much did 'EndGraph' improve preprocessing performance compared to the group of 'LFGraph', 'PowerLyra', 'PowerGraph', and 'D-Galois'?''}
\end{itemize}

\end{document}